\documentclass[aps,prl,twocolumn,groupedaddress]{revtex4-2}

\usepackage[colorlinks,urlcolor=blue, citecolor=blue,linkcolor=blue]{hyperref}
\usepackage{verbatim} 
\usepackage{graphicx}
\usepackage{amsmath}
\usepackage{xcolor}

\usepackage[sort&compress]{natbib}

\begin{document}
\title{Measurement of the $\boldsymbol{e^{+}e^{-}\to K^+K^-\psi(2S)}$ Cross Section at Center-of-Mass Energies from 4.699 to 4.951 GeV and Search for $\boldsymbol{Z_{cs}^{\pm}}$ in the $\boldsymbol{Z_{cs}^\pm\to K^\pm\psi(2S)}$ Decay}

\author{
\begin{small}
\begin{center}
M.~Ablikim$^{1}$, M.~N.~Achasov$^{4,c}$, P.~Adlarson$^{76}$, O.~Afedulidis$^{3}$, X.~C.~Ai$^{81}$, R.~Aliberti$^{35}$, A.~Amoroso$^{75A,75C}$, Q.~An$^{72,58,a}$, Y.~Bai$^{57}$, O.~Bakina$^{36}$
% ...
\\
\vspace{0.2cm}
(BESIII Collaboration)\\
\vspace{0.2cm} {\it
$^{1}$ Institute of High Energy Physics, Beijing 100049, People's Republic of China\\
% ...
$^{3}$ Bochum  Ruhr-University, D-44780 Bochum, Germany\\
$^{4}$ Budker Institute of Nuclear Physics SB RAS (BINP), Novosibirsk 630090, Russia\\
% ...
$^{35}$ Johannes Gutenberg University of Mainz, Johann-Joachim-Becher-Weg 45, D-55099 Mainz, Germany\\
$^{36}$ Joint Institute for Nuclear Research, 141980 Dubna, Moscow region, Russia\\
% ...
$^{57}$ Southeast University, Nanjing 211100, People's Republic of China\\
$^{58}$ State Key Laboratory of Particle Detection and Electronics, Beijing 100049, Hefei 230026, People's Republic of China\\
% ...
$^{72}$ University of Science and Technology of China, Hefei 230026, People's Republic of China\\
% ...
$^{75}$ University of Turin and INFN, (A)University of Turin, I-10125, Turin, Italy; (B)University of Eastern Piedmont, I-15121, Alessandria, Italy; (C)INFN, I-10125, Turin, Italy\\
$^{76}$ Uppsala University, Box 516, SE-75120 Uppsala, Sweden\\
% ...
$^{81}$ Zhengzhou University, Zhengzhou 450001, People's Republic of China\\

\vspace{0.2cm}
$^{a}$ Deceased\\
% ...
$^{c}$ Also at the Novosibirsk State University, Novosibirsk, 630090, Russia\\
% ...

}

\end{center}
\vspace{0.4cm}
\end{small}
}


\begin{abstract}
We perform the first investigation of the process $e^{+}e^{-}\to K^+K^-\psi(2S)$ and report its Born cross sections over a range of center-of-mass energies from 4.699 to 4.951~GeV.
The measurements are carried out using several partial reconstruction techniques 
using data samples collected by the BESIII detector with a total integrated luminosity of 2.5~fb$^{-1}$.
We search for new tetraquark candidates $Z_{cs}^\pm$ in the decays $Z_{cs}^\pm\to K^\pm\psi(2S)$.
No significant $Z_{cs}^\pm$ signals are observed.

\end{abstract}.

\maketitle

At the frontier of Quantum Chromodynamics~(QCD), exotic hadrons containing heavy quarks have 
been the subject of much experimental and theoretical effort.  Interest was originally stimulated two decades ago with the discovery of the first exotic~(non-$q\bar{q}$) candidate, the $X(3872)$, by Belle in 2003~\cite{Belle:2003nnu}.
Subsequently, additional exotic candidates involving a charm-anticharm pair, \textit{e.g.}, the $Y(4260)$~\cite{babar-y4260-1} and $Z_c(3900)$~\cite{ref_tetra_bes1,ref_tetra_belle}, were also experimentally established.  Interpreting their nature is a high priority for both experiment and theory.

Various experiments have 
measured a series of $e^+e^-$ cross sections to both hidden-charm and open-charm final states,
\textit{e.g.}, $e^+e^-\to\pi^+\pi^-J/\psi$~\cite{bes-ppjpsi}, 
$\pi^+\pi^- h_c$~\cite{bes-pphc},
$\pi^+\pi^- \psi(2S)$~\cite{bes-pppsi},
and $D^{(*)}D^{(*)} (\pi)$~\cite{bes-open-xp,
bes-open-wb,bes-open-ddp}, and have
observed candidate vector states in the center-of-mass energy~($\sqrt{s}$) depenedence of those cross sections that do not fit within the conventional 
charmonium spectrum.
Moreover, processes with strange mesons in the final state, 
such as $e^+e^-\to K\bar{K}J/\psi$~\cite{belle-kkjpsi,
bes-kkjpsi1,bes-kkjpsi2,bes-kkjpsi3},
have also been studied. 
BESIII observed a new structure in the $\sqrt{s}$-dependence of 
the $e^+e^-\to K^+K^-J/\psi$ cross section~\cite{bes-kkjpsi3} at a mass of 4.710 GeV, called
the $Y(4710)$.
This structure is one of the heaviest vector charmonium-like states observed to date. 
BESIII also measured the $\sqrt{s}$-dependence of the cross section for the process
$e^+e^-\to D_s^{*+}D_s^{*-}$~\cite{bes-dds}, and found a structure around 4.79 GeV, with a mass that is 
distinct from the $Y(4710)$. 
Further studies are needed to clarify the nature of these two structures in 
processes containing strange mesons. 
A study of the process $e^+e^-\to K^+K^-\psi(2S)$ is an important extension of previous efforts,
and the $e^+e^-$ annihilation datasets collected above the $e^+e^-\to K^+K^-\psi(2S)$ threshold at BESIII
make this possible.

Unlike the $X$ and $Y$ states, the isovector $Z$ states must be exotic
since they contain at least a light quark and antiquark in addition to an isosinglet heavy quark–antiquark pair.
In 2013, the tetraquark candidate $Z_c(3900)^\pm$
was observed decaying to $\pi^\pm J/\psi$ in the 
$\pi^+\pi^- J/\psi$ system
by the BESIII and Belle experiments~\cite{ref_tetra_bes1,ref_tetra_belle}.
Charged structures have also 
been observed in the $\pi^{\pm}\psi(2S)$ invariant mass spectrum in
both the $B\to K\pi^\pm\psi(2S)$~\cite{b_psipi1} and
$e^+e^-\to\pi^+\pi^-\psi(2S)$~\cite{bes_psipi} reaction channels.
Several tetraquark candidates 
have been observed in both hidden- and open-charm processes.
BESIII reported 
the observation of a charged 
open-strange hidden-charm structure $Z_{cs}(3985)$ in the $K^+$ 
recoil-mass spectra in the
$e^+e^- \rightarrow K^+(D_s^-D^{*0}+D_s^{*-}D^0)$ process~\cite{ref_tetra_zcs} with a
mass and width of
$M=3992.2\pm1.7\pm1.6$~MeV/$c^2$ and $\Gamma=7.7^{+4.1}_{-3.8}\pm4.3$~MeV.
This was the first observation of a tetraquark candidate involving both 
strange and charm quarks.
Soon after, LHCb reported observations of the
 $Z_{cs}(4000)$ and $Z_{cs}(4220)$ decaying to $J/\psi K^+$ via the reaction $B^+\to J/\psi\phi K^+$~\cite{ref_tetra_zcs_lhcb}.
The mass and width of the $Z_{cs}(4000)$
were found to be $M=4003\pm6^{+4}_{-14}$~MeV/$c^2$ and $\Gamma=131\pm15\pm26$~MeV, and 
for the $Z_{cs}(4220)$, 
$M=4216\pm{24}^{+43}_{-30}$~MeV/$c^2$ and $\Gamma=233\pm52^{+97}_{-73}$~MeV.
The $Z_{cs}(3985)$ from BESIII 
and the $Z_{cs}(4000)$ from LHCb have similar masses 
but quite different widths.
According to calculations in 
the hadro-charmonium picture~\cite{zcs_theory,zcs_ar}, the
$Z_{cs}(3985)$ and $Z_{cs}(4000)$
can be assigned to $\psi(2S)\otimes K$ hadro-charmonia,
while the $Z_{cs}(4220)$ 
could be assigned to a 
$\psi(2S)\otimes K^*$ or $\chi_{c1}(2P)\otimes K^*$ state. In addition,
BESIII searched for $Z_{cs}\to KJ/\psi$ in the process 
 $e^+e^-\to K^+K^- J/\psi$,
but no significant signals 
were observed~\cite{bes-kkjpsi3}. As a natural extension, 
a search for the $Z_{cs}$ in the decay $Z_{cs}\to K^\pm\psi(2S)$ is very 
interesting.

In this Letter, 
we report the first measurement of the $e^{+}e^{-}\to K^+K^-\psi(2S)$ Born cross sections ($\sigma$)
at $\sqrt{s}$ from 4.669 to 4.951 GeV
using $2.5~\rm fb^{-1}$ of $e^+e^-$ annihilation data collected by the BESIII detector.
To investigate intermediate $Y$ states that could be produced through the reaction $e^{+}e^{-}\to Y \to K^+K^-\psi(2S)$, the $\sqrt{s}$-dependent ratio between
$\sigma(e^{+}e^{-}\to K^+K^-\psi(2S))$ and $\sigma(e^{+}e^{-}\to K^+K^- J/\psi)$ is also provided.
Taking advantage of the largest signal yield at 4.843 GeV, 
we search for new tetraquark candidates $Z_{cs}^\pm$ in the $Z_{cs}^\pm\to K^\pm\psi(2S)$ decay 
channels.

The BESIII detector is described in Ref.~\cite{BES}.
The Monte Carlo (MC) samples are generated with {\sc kkmc}~\cite{KKMC} in conjunction with {\sc evtgen}~\cite{evtgen}. 
The detector simulation is based on {\sc geant4}~\cite{geant}.
The inclusive MC samples, which include open-charm hadronic processes, continuum processes, and the effects due to initial-state-radiation (ISR), are produced with ten times the data luminosity to study the backgrounds. 
The signal MC samples, 
$e^{+}e^{-}\to K^+K^-\psi(2S)$, $\psi(2S)\to J/\psi+(\pi^+\pi^-,\pi^0\pi^0,\eta,\pi^0,\gamma\gamma)$ or $ \ell^+\ell^-$,
$J/\psi\to \ell^+\ell^-$ ($\ell=e,\mu$),
are generated to determine the detection efficiencies.
The criteria of charged track selection and kaon identification are the same as 
those in Ref.~\cite{bes-kkjpsi3}.
To reconstruct the $J/\psi$ or $\psi(2S)$ in their leptonic decays,
two charged particles with momenta greater than 1.0 GeV/$c$ and opposite charges
are identified as the lepton pair from the $J/\psi$ or $\psi(2S)$ decay.
Electrons and muons are discriminated by requiring
their deposited energies in the electromagnetic calorimeter (EMC) to be 
greater than 0.8~GeV and less than 0.4~GeV, respectively.

A partial reconstruction technique is used to improve the reconstruction efficiency in the measurement.
We take advantage of two aspects of the signal signature.
First, the $\psi(2S)$ decays dominantly through channels with a $J/\psi$ in the final state,
including $\psi(2S)\to J/\psi+(\pi^+\pi^-,\pi^0\pi^0,\eta,\pi^0,\gamma\gamma)$~\cite{PDG2022}.
Furthermore, since the $\sqrt{s}$ range is close to 
the threshold for $e^{+}e^{-}\to K^+K^-\psi(2S)$ production, the primary kaons have low momentum and
low detection efficiency. 
Based on these features, four approaches are developed to 
reconstruct the $e^{+}e^{-}\to K^+K^-\psi(2S)$ signals.
The selection criteria of the four approaches are orthogonal and do not contain events in common.

In the first approach~(i), we tag the $K^+$, the $K^-$, and the $J/\psi$ from the $\psi(2S)$ decay, and then identify the signal $\psi(2S)$ using the mass recoiling against the 
$K^+K^-$ system ($RM(K^+K^-)$). 
This approach aims to reconstruct signals using all $\psi(2S)$ decay channels that contain
a $J/\psi$ in the final state.
The $J/\psi$ mass window is set to be $3.05 < M(\ell^+\ell^-) < 3.15$~GeV/$c^2$, where 
$M(\ell^+\ell^-)$ is the lepton pair invariant mass.
The selection criteria are optimized by maximizing $S/\sqrt{S+B}$, where $S$ is
the number of signal events and $B$ is the number of background events, 
estimated according to the inclusive MC samples.
To suppress backgrounds caused by $\mu/\pi$ misidentification, at least one muon
from $J/\psi\to \mu^+\mu^-$ needs to penetrate more than three layers 
of the muon chamber (MUC).

In the second approach~(ii), we only tag one kaon (the $K^+$ or $K^-$) by requiring there be exactly one charged kaon detected 
and then reconstruct the $\psi(2S)$ through the 
$\psi(2S)\to\pi^+\pi^- J/\psi$ decay. A one-constraint kinematic fit constraining the mass of the 
missing kaon is further applied and is used to extract its four-momentum.
The resulting $\chi^2_{1\rm C}$ is required 
to be less than 50.
The requirement on the
$J/\psi$ mass window is the same as that in (i).
This method is designed to recover those signal candidates with only one reconstructed kaon.

In the third approach~(iii), we tag the $K^+$, the $K^-$, and the $\psi(2S)$ in the decay $\psi(2S)\to \ell^+\ell^- (\ell=e,\mu)$.
The $\psi(2S)$ mass window is required to be $3.631 < M(\ell^+\ell^-) < 3.726$~GeV/$c^2$.

In the fourth approach~(iv), we tag only one kaon (the $K^+$ or $K^-$), analogous to (ii), but the $\psi(2S)$ is reconstructed through
$\psi(2S)\to \ell^+\ell^-$.  
The $\chi^2_{1\rm C}$ of the kinematic fit to the missing kaon mass is required to be less than 15.
To suppress Bhabha background in the $\psi(2S)\to e^+e^-$ channel, events with $e^+$
 and $e^-$ polar angles in the region $\rm cos(\theta_{e^+})>0.85$ and $\rm cos(\theta_{e^-})<-0.85$
are vetoed.

\begin{figure}
\includegraphics[scale=0.4]{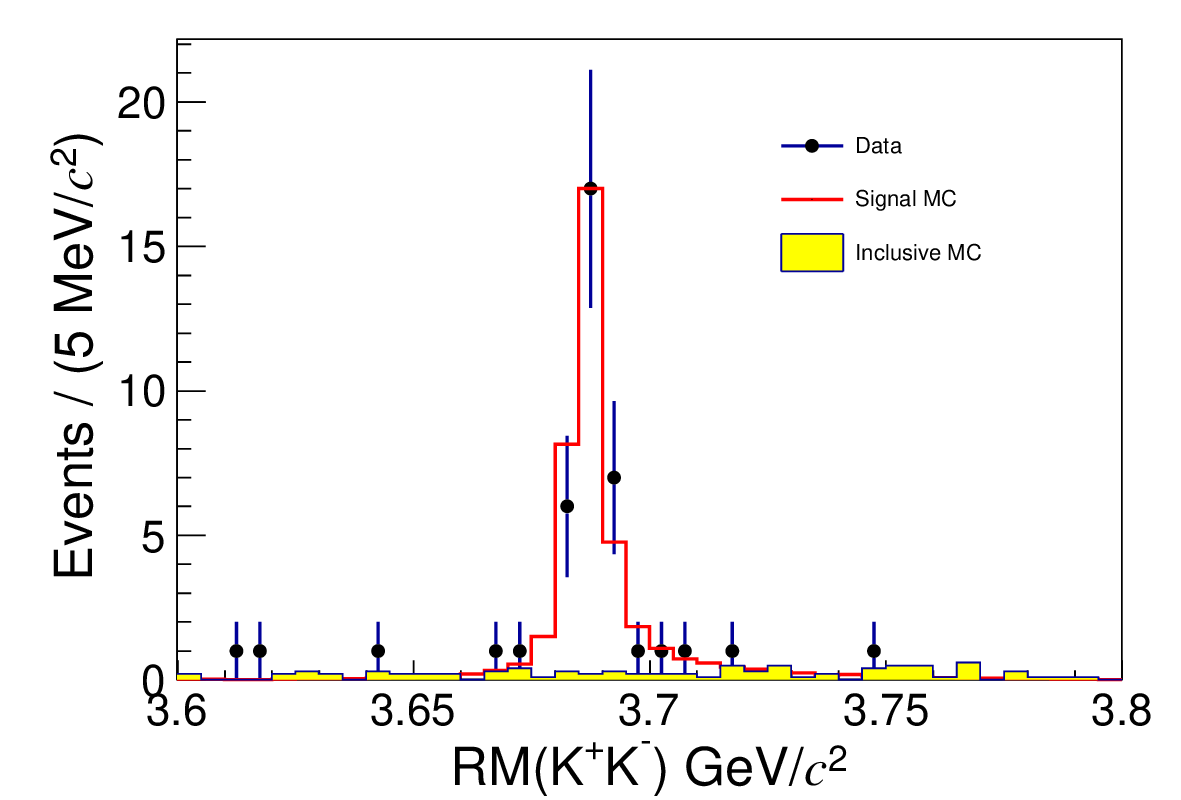}
\caption{Distribution of $RM(K^+K^-)$ at $\sqrt{s}=4.843$~GeV.
The points with error bars are data, the red histogram is for the signal MC events, 
and the yellow filled histogram is for the inclusive MC events.}
\label{fig:fit_rmkk_4843}
\end{figure}

Figure~\ref{fig:fit_rmkk_4843} shows 
the total $RM(K^+K^-)$  distribution using selected events from all 
four approaches at $\sqrt{s}=4.843$ GeV.
A distinct $\psi(2S)$ signal peak is evident.
According to a study of the inclusive MC samples,
there are no peaking backgrounds.
The signal yield at each energy point is 
obtained by counting events 
in the $\psi(2S)$ signal region  
[3.67, 3.71] GeV/$c^2$, 
which covers around $\pm3\sigma$ of the signal shape 
according to the 
signal MC distribution. 
The background in 
the signal region is estimated  using sidebands that are 
two times wider than the signal region.  For $\sqrt{s}>4.8$ GeV, the sidebands are
[3.61,3.65] and [3.73, 3.77] GeV/$c^2$, while for $\sqrt{s}<4.8$ GeV, due to the limited available phase space, only the lower energy sideband is used.
Assuming the observed 
events in the signal and sideband regions follow Poisson distributions, 
a likelihood defined as $\mathcal{L}(x,y|s,b,\tau)=Pois(x|s+\tau b)Pois(y|b)$ is used to 
extract the signal yield, the statistical uncertainties, and the signal significance.
Here, $x(s)$ and $y(b)$ correspond to the observed (expected) yields in the signal
and sideband regions, respectively, and $\tau$ is the ratio of the width of the signal region 
to that of the sideband.
The maximum likelihood (ML) $\mathcal{L}_{\textrm{max}}$ 
is obtained by scanning $s$ and $b$,
where the corresponding $s$ value is taken as the 
signal yield.
The errors are the difference between the $s$ with a likelihood of
$e^{-0.5}\mathcal{L}_{\textrm{max}}$ and $\mathcal{L}_{\textrm{max}}$, respectively.
The significance is estimated by comparing 
the difference of log ML values 
$\Delta(-2\rm{ln}\mathcal{L}_{\textrm{max}})$
by setting $s$ to be zero and nonzero.
All results are listed in the Supplemental Material~\cite{supp}.

The Born cross section of $e^+e^-\to K^+K^-\psi(2S)$
is calculated as
\begin{equation}\label{equ:cor}
\sigma^{\rm{B}}=\frac{N_\mathrm{s}}{\mathcal{L}_{\rm{int}}\epsilon_{r}(1+\delta)\frac{1}{|1-\Pi|^{2}}},
\end{equation}
where
$N_\mathrm{s}$ is the number of 
signal events and
$\mathcal{L}_{\rm{int}}$ is the 
integrated luminosity.
The efficiency $\epsilon_{r}$ is $\mathcal{B}(\ell\ell)(\epsilon_{\rm iv}+\epsilon_{\rm iii})+[\mathcal{B}(\pi^+\pi^-J/\psi)\epsilon_{\rm ii}+\mathcal{B}(XJ/\psi)\epsilon_{\rm i}]\mathcal{B}(J/\psi\to\ell\ell)$, 
where $\mathcal{B}(\rm \ell\ell)$ is 
the sum of the branching fractions for $\mathcal{B}(\psi(2S)\to ee)$ and 
$\mathcal{B}(\psi(2S)\to\mu\mu)$, 
$\mathcal{B}(J/\psi\to\rm \ell\ell)$ is 
the combined branching fraction for $\mathcal{B}(J/\psi\to ee)$ and 
$\mathcal{B}(J/\psi\to\mu\mu)$, 
$\mathcal{B}(XJ/\psi)$ is the total branching fraction 
of all $\psi(2S)$ decays that contain a $J/\psi$, and
$\mathcal{B}(\pi^+\pi^-J/\psi)$ is the branching fraction
of $J/\psi\to \pi^+\pi^-J/\psi$.
All of these branching fractions are taken 
from the Particle Data Group~\cite{PDG2022}.
In addition, $\epsilon_{\rm i}$, $\epsilon_{\rm ii}$,
$\epsilon_{\rm iii}$, and $\epsilon_{\rm iv}$
are the average efficiencies of the electron and muon channels for the four reconstruction 
approaches, respectively,
and $(1+\delta)$ is the radiative correction factor 
obtained by a QED calculation~\cite{QED-Delt}.
A Breit-Wigner (BW) function is used to describe 
the lineshape of the observed cross section
and extract the 
ISR corrected efficiencies and $(1+\delta)$ by iterating 
the input lineshape until convergence according to 
the method in Ref.~\cite{isr_med}. 
The $\frac{1}{|1-\Pi|^{2}}=1.055$ is the vacuum polarization 
factor taken from QED with an accuracy of 0.05\%~\cite{Vacuum-Delt}.
The measured Born cross sections are listed in the Supplemental Material~\cite{supp}.

\begin{figure}
\includegraphics[scale=0.4]{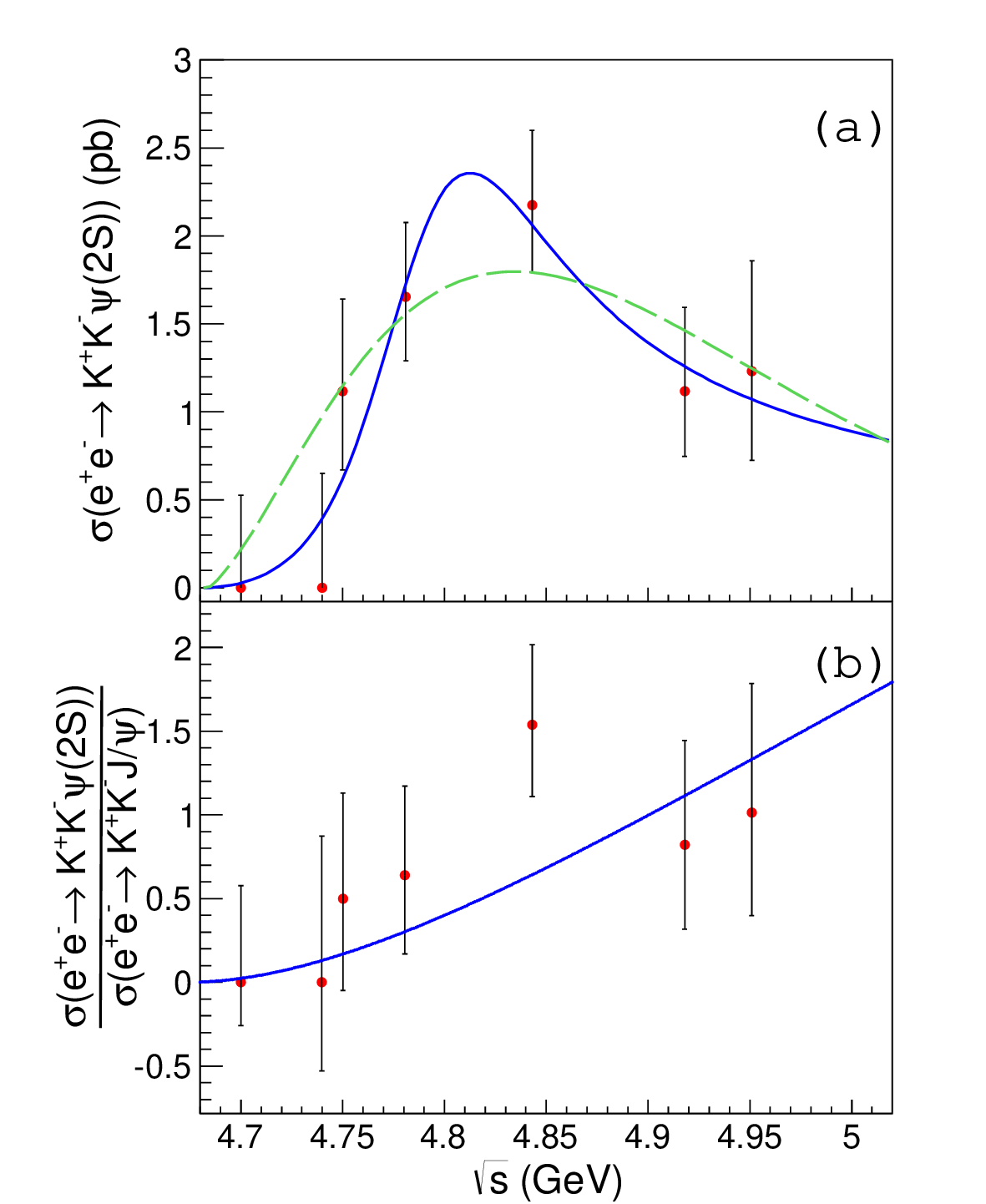}
\caption{ 
The $\sqrt{s}$-dependent cross section of $e^+e^-\to K^+K^-\psi(2S)$
is shown in (a), while the ratio $\sigma(K^+K^-\psi(2S))$/$\sigma(K^+K^-J/\psi)$
in (b).
The error bars are statistical only. 
In (a), the solid curve is the fit using a single BW function,
and the dashed curve denotes the exponential function.
In (b), the solid curve is the ratio of
the phase space of the three-body $K^+K^-\psi(2S)$ to $K^+K^-J/\psi$ reactions.
}
\label{fig_cr_ra}
\end{figure}

The dressed cross section ($\sigma^{\rm B}/|1-\Pi|^{2}$)
is shown in Fig.~\ref{fig_cr_ra} (a) as a function of $\sqrt{s}$.
By assuming the observed $e^+e^-\to K^+K^-\psi(2S)$
signals are from a vector resonance $Y$ decay, 
a phase space modified BW function is fit to the energy-dependent dressed cross section using
$\sigma^{\textrm{dressed}}=|BW(\sqrt{s})|^2$, with $BW(\sqrt{s})$ defined as   
\begin{equation}\small
BW(s)=\frac{M}{\sqrt{s}}\cdot\frac{\sqrt{12\pi\Gamma_{\textrm{tot}}\Gamma_{ee}\mathcal{B}_{Y\to K^+K^-\psi(2S)}}}{s-M^2+iM\Gamma_{\textrm{tot}}}\cdot\sqrt{\frac{\Phi(\sqrt{s})}{\Phi(M)}},
\end{equation}
where $M$, $\Gamma_{\textrm{tot}}$, and $\Gamma_{ee}$
are the mass, total width, and di-electron partial width of the resonance $Y$, respectively;
$\mathcal{B}_{Y\to K^+K^-\psi(2S)}$ is the branching fraction of 
$Y\to K^+K^-\psi(2S)$; and
the phase space of the three-body decay is 
$\Phi(\sqrt{s})=\iint\frac{1}{(2\pi)^332(\sqrt{s})^3}dm^2_{23}dm^2_{12}$.

A ML fit is performed to extract 
the resonance parameters.  
The fit results are $M=4787.7\pm17.7$ MeV/$c^2$,
$\Gamma=110.3\pm33.9$ MeV, and
$\Gamma_{ee}\mathcal{B}_{Y\to K^+K^-\psi(2S)}=0.13\pm0.02$ eV.
The ML value is $-\rm{ln}\mathcal L=-30.9$.
Alternatively, 
the signals could be produced from the decay of an established resonance, e.g. 
$Y(4710)$~\cite{bes-kkjpsi3}, and 
the continuum process. To examine this, we use an 
exponential function as used in Refs.~\cite{bes_ppj,babar_ppj}, 
$\sigma(\sqrt{s})=p_1\cdot\Phi(\sqrt{s})e^{p_0(\sqrt{s}-M_{\textrm{th}})}$,
to describe the line shape,
where $p_0$ and $p_1$ are the free parameters, and
$M_{\textrm{th}}=2m_{K^{\pm}}+m_{\psi(2S)}$.
The ML of the fit is $-\rm{ln}\mathcal{L}=-29.5$, 
similar to the previous approach.

In addition, we provide the $\sqrt{s}$-dependent ratio
of $\sigma(e^+e^-\to K^+K^-\psi(2S))$ to $\sigma(e^+e^-\to K^+K^-J/\psi)$
as shown in Fig.~\ref{fig_cr_ra}(b), where
the $\sqrt{s}$-dependent Born cross sections $\sigma(e^+e^-\to K^+K^-J/\psi)$ are taken from the  
BESIII measurement in Ref.~\cite{bes-kkjpsi3}.
If the reactions $e^+e^-\to K^+K^-\psi(2S)$ and $e^+e^-\to K^+K^-J/\psi$
proceed through the same mechanisms,
the $\sqrt{s}$-dependent ratio is likely to
indicate a similar trend to that of 
phase space, shown as the solid curve in Fig.~\ref{fig_cr_ra}(b).
The measured ratio at $\sqrt{s}=$ 4.843 GeV has about a
$2\sigma$ statistical deviation from that of phase space, which could indicate that a distinct production mechanism for
$e^+e^-\to K^+K^-\psi(2S)$ may exist.

The systematic uncertainties for the Born cross section measurement mainly 
originate from the detection efficiency,
the ISR correction factor, the integrated luminosity, and the input branching 
fractions.
The sources of the uncertainty from the detection efficiencies include the tracking, 
particle identification, the kinematic fit, the $J/\psi$/$\psi(2S)$ mass window, 
the muon identification with MUC, and the signal generation model. 
The systematic uncertainty due to tracking and particle identification 
is 1.0\% for each track according to studies of the control samples 
$e^+e^-\to\pi^+\pi^-J/\psi$~\cite{trk} and
$J/\psi\to K_{S}^0 K^\pm\pi^\mp$~~\cite{pid}.
The uncertainties caused by the kinematic fit, 
the $J/\psi$ mass selection, and the muon selection with the MUC are studied 
with a control sample of $e^+e^-\to\pi^+\pi^- \psi(2S)$. 
To estimate the systematic uncertainty caused 
by the generator model, we produce the MC samples of the processes
$e^+e^-\to f \psi(2S)\to K^+K^-\psi(2S)$ ($f$ =
$f_2(1270)$, $f_0(1370)$, $f_0(1500)$, $f_2^\prime(1525)$) or $e^+e^-\to K^{\pm}Z_{cs}^{\pm}\to K^{+}K^-\psi(2S)$. 
The efficiency difference compared to the nominal value is  4.3\% and is taken as 
the systematic uncertainty.
The uncertainty due to the ISR correction factor
is estimated by replacing the BW function in the MC generation with the 
exponential function applied in the fit scheme. 
The integrated luminosity is measured with the 
Bhabha scattering process with an uncertainty 
of 1.0\%~\cite{lum}. The total systematic uncertainty at each energy point is obtained 
by adding all these systematic uncertainties in quadrature. 
The systematic uncertainties discussed above are summarized in the Supplemental Material~\cite{supp}.

\begin{figure}[!h]
\begin{center}
\includegraphics[scale=0.4]{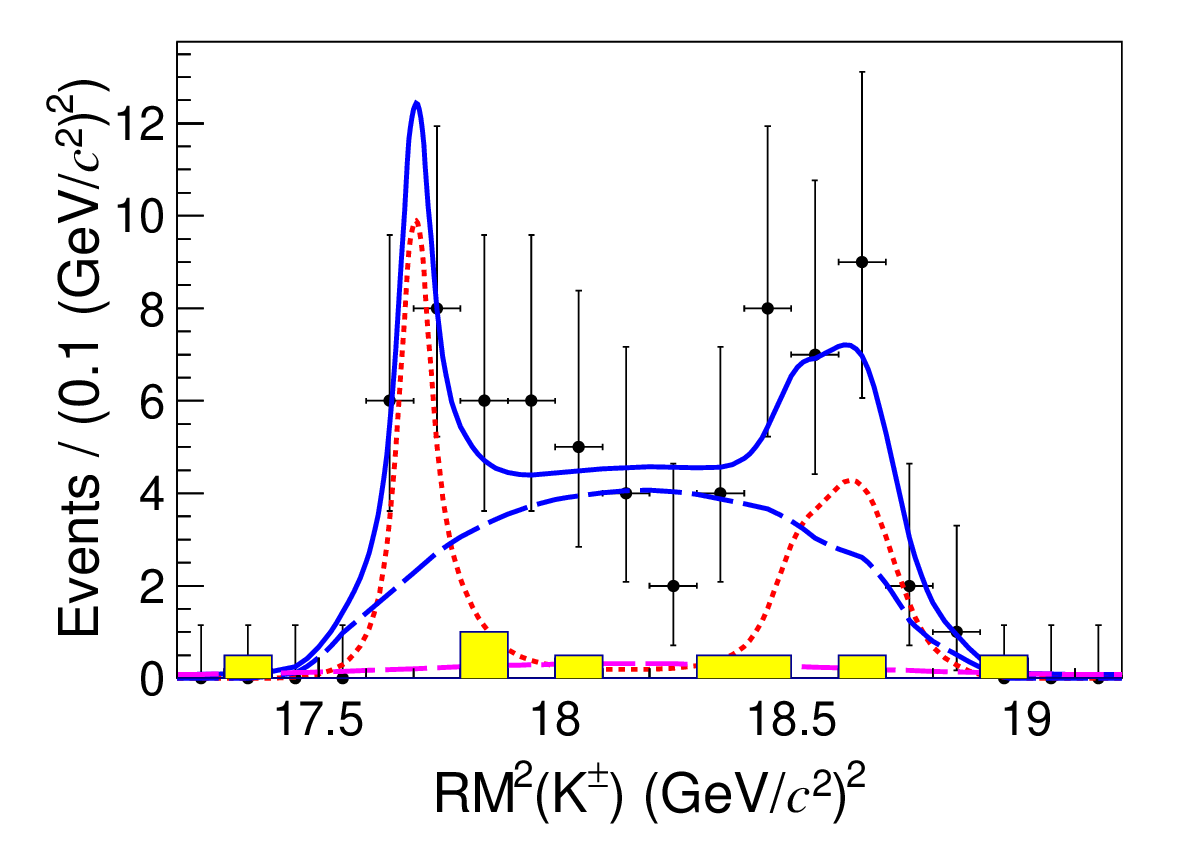}
\includegraphics[scale=0.4]{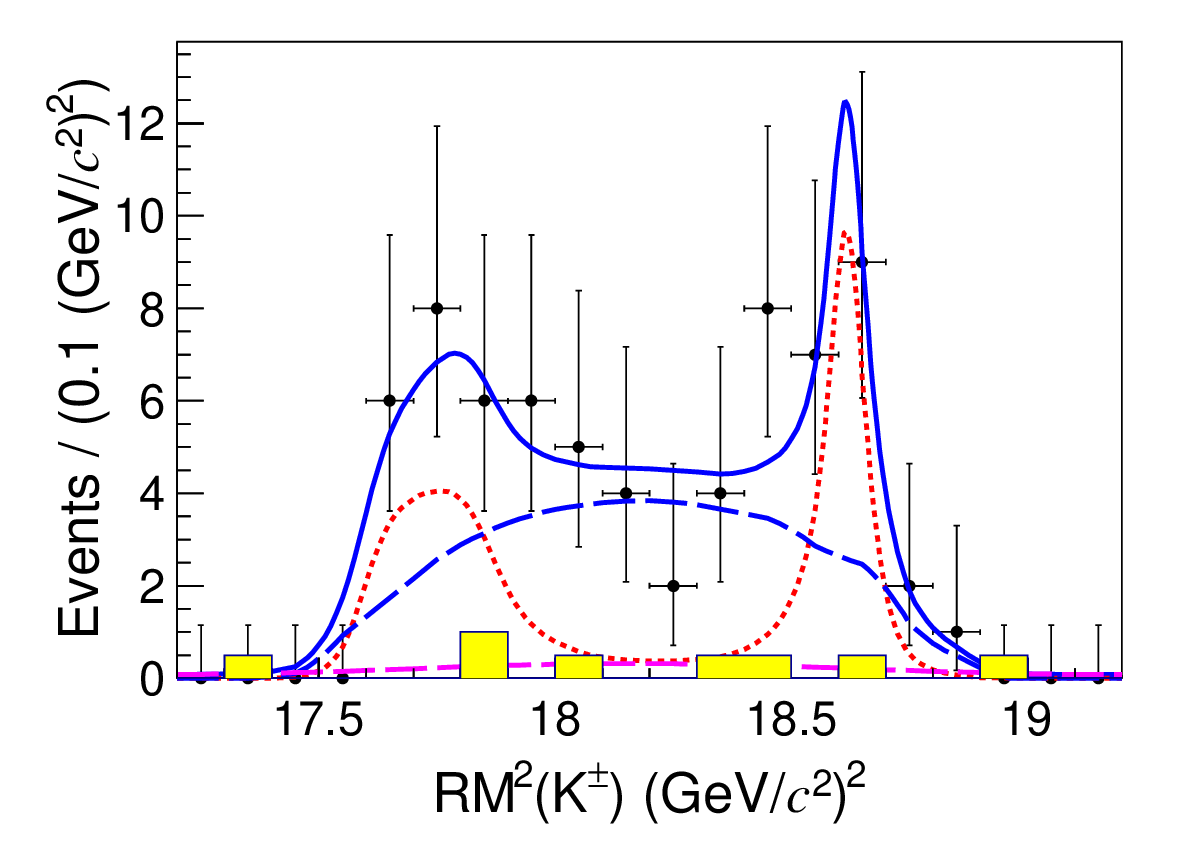}
\caption{
Distribution of $RM^2(K^\pm)$.
The top plot shows the
results from Fit I, and the lower one shows the Fit II results.
The dots with error bars are data,
the solid blue lines are the total fit results,
the red dotted lines represent the signal, 
the blue dashed line indicate the $e^+e^-\to K^+K^-\psi(2S)$ non-resonant contributions,
and the pink dashed lines show the non-$K^+K^-\psi(2S)$ background. 
The sideband distributions are shown by the yellow filled histograms.
}
\label{fig:fig_rmk2_fit1}
\end{center}
\end{figure}

Since the data sample at $\sqrt{s}=4.843$ GeV gives the largest 
$e^+e^-\to K^+K^-\psi(2S)$ signal yield, we use it to 
search for intermediate states in the $K^+K^-\psi(2S)$ system.
Figure~\ref{fig:fig_rmk2_fit1} shows the invariant mass squared distribution of the system recoiling against the $K^{\pm}$, $RM^2(K^\pm)$.

\begin{figure}
\includegraphics[scale=0.4]{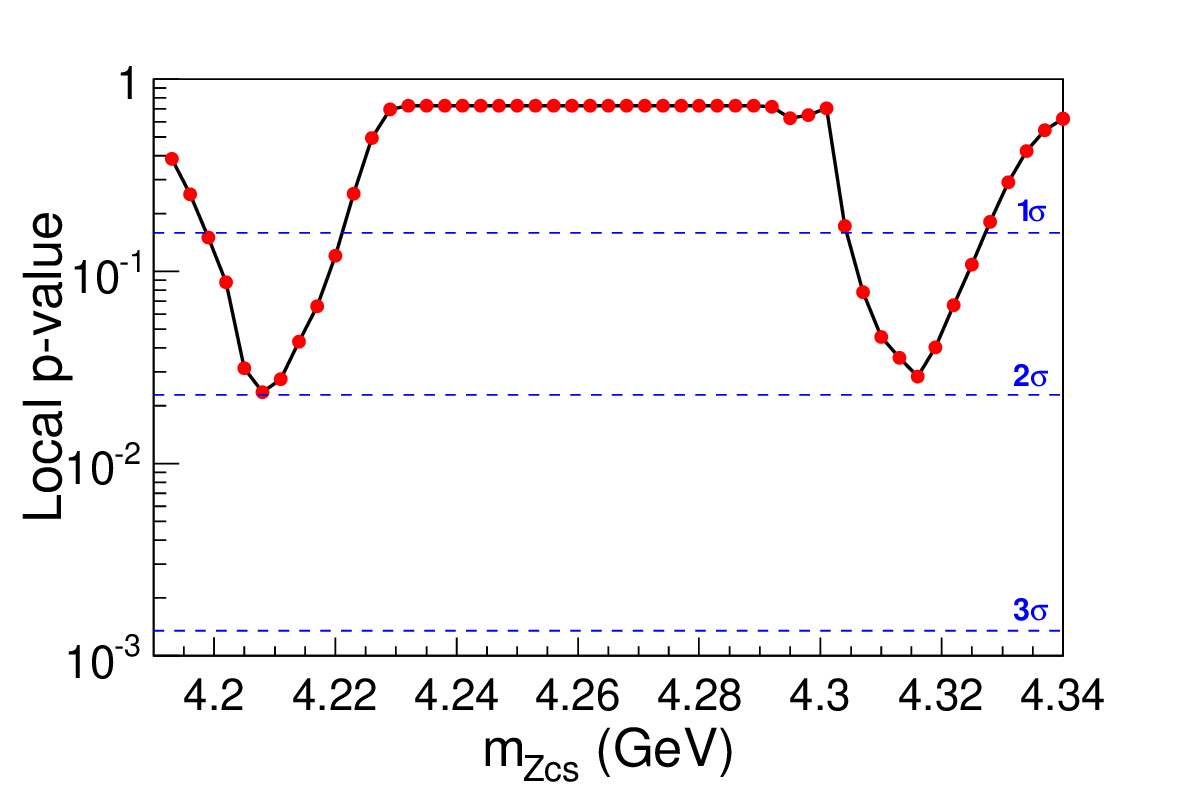}
\caption{The local p-values as a function of the $Z_{cs}$ mass.}
\label{fig:localp}
\end{figure}

We perform a simultaneous 
fit to the $RM^2(K^+)$ and $RM^2(K^-)$ spectra
to extract the $Z_{cs}^{\pm}$ signal yield.
The fit function consists of three components: the $Z_{cs}^{\pm}$
signal, contributions from  the $e^+e^-\to K^+K^-\psi(2S)$ non-resonance process, and 
non-($K^+K^-\psi(2S)$) backgrounds.
The $Z_{cs}^{\pm}$ signal shape is decribed with 
$f(X) = \sigma(X) \otimes \int PHSP\cdot(|\frac{1}{X-M^2-iM\Gamma}|^2+|\frac{1}{Y-M^2-iM\Gamma}|^2) dY$,
where $X,~Y$ are $RM^2(K^+)$ and $RM^2(K^-)$, respectively, and $M$ and 
$\Gamma$ are the $Z_{cs}$ mass and width.
$PHSP$ denotes the two-dimensional distribution of $RM^2(K^+)$
versus $RM^2(K^-)$ obtained 
with the $e^+e^-\to K^+K^-\psi(2S)$ phase space MC sample, and
$\sigma(X)$ is the resolution 
of $RM^2(K^\pm)$.  
The shape of the $e^+e^-\to K^+K^-\psi(2S)$ non-resonance process
is obtained from the $e^+e^-\to K^+K^-\psi(2S)$ phase space MC sample. The contribution of
the non-$K^+K^-\psi(2S)$ background is estimated by
the $\psi(2S)$ sidebands in data.
In the fit,  
the mass and width of the resonance, the magnitudes of 
the $Z_{cs}$ signal and the non-resonance $e^+e^-\to K^+K^-\psi(2S)$ 
process are free parameters. The interference between different 
processes are ignored due to the limited statistics.
To localize the position of the $Z_{cs}$ signal, 
a series of fits are implemented by scanning the $Z_{cs}$ mass in 
the physical mass region. The local p-value is obtained 
by comparing the likelihood to that of the 
background-only null hypothesis. 
Figure~\ref{fig:localp} shows the 
local p-values as a function of the $Z_{cs}$ mass.
The masses around 4.205 GeV and 4.315 GeV
give the minimum local p-values.

We perform two fits to the $RM^2(K^\pm)$ distribution.  In Fit~I,  the mass of the $Z_{cs}$ is assumed to be around 4.205~GeV with a reflection 
at higher mass.  In Fit~II,  the mass of the $Z_{cs}$ is around 4.315~GeV 
with a reflection at lower mass.
Figure~\ref{fig:fig_rmk2_fit1}
shows the fit results.
Fit I gives a mass and width for the $Z_{cs}^{\pm}$ of $M=4208.4\pm3.1$ MeV/$c^2$ and 
$\Gamma=6.1\pm5.7$ MeV, with a global significance including the look-elsewhere effect
of 1.2$\sigma$. 
For Fit II, $M=4316.0\pm2.7$ MeV/$c^2$ and 
$\Gamma=9.0\pm8.6$ MeV, 
with a global significance 
of 1.1$\sigma$. The uncertainties here are statistical only.

In summary, we investigate the process $e^{+}e^{-}\to K^+K^-\psi(2S)$ 
using data samples 
at $\sqrt{s}$ from 4.699 to 4.951 GeV 
collected by the BESIII detector at the BEPCII collider.
We report the first measurement of the $e^{+}e^{-}\to K^+K^-\psi(2S)$ Born cross sections.
The $\sqrt{s}$-dependence can be well-described as the decay of a single vector resonance with or without a superimposed continuum process.  However, an empirical non-resonant function produces similar fit quality.  
Furthermore, the $\sqrt{s}$-dependent ratio of 
$\sigma(e^{+}e^{-}\to K^+K^-\psi(2S))$ to $\sigma(e^{+}e^{-}\to K^+K^- J/\psi)$ is provided using 
values taken from our measurement in Ref.~\cite{bes-kkjpsi3}.
At $\sqrt{s}$ = 4.843~GeV,
a deviation of about $2\sigma$ with respect to the ratio of their phase spaces is found, which may imply
a new resonance with hidden strangeness to produce the $K^+K^-\psi(2S)$ signals.
We search for new tetraquark candidates $Z_{cs}^\pm$ in the $Z_{cs}^\pm\to K^\pm\psi(2S)$ decay through the 
study of the observed $e^{+}e^{-}\to K^+K^-\psi(2S)$ signals.
The simultaneous fit to the $RM^2(K^+)$ 
and $RM^2(K^-)$ spectra 
gives two best fit results with the $Z_{cs}^{\pm}$ masses around 4.208~GeV/$c^2$ and 4.315~GeV/$c^2$, respectively. A mass of 4.208~GeV/$c^2$ is in the vicinity of the $Z_{cs}(4220)$ reported by LHCb~\cite{ref_tetra_zcs_lhcb}.
These measurements add to our knowledge of exotic hadrons with strangeness, and provide inspiration  
for new research directions in both the theoretical and experimental sectors.

\textbf{Acknowledgement}

The BESIII Collaboration thanks the staff of BEPCII and the IHEP computing center for their strong support. This work is supported in part by National Key R\&D Program of China under Contracts Nos. 2020YFA0406300, 2020YFA0406400, 2023YFA1606000; National Natural Science Foundation of China (NSFC) under Contracts Nos. 11635010, 11735014, 11935015, 11935016, 11935018, 12025502, 12035009, 12035013, 12061131003, 12192260, 12192261, 12192262, 12192263, 12192264, 12192265, 12221005, 12225509, 12235017, 12361141819; the Chinese Academy of Sciences (CAS) Large-Scale Scientific Facility Program; the CAS Center for Excellence in Particle Physics (CCEPP); Joint Large-Scale Scientific Facility Funds of the NSFC and CAS under Contract No. U1832207; 100 Talents Program of CAS; The Institute of Nuclear and Particle Physics (INPAC) and Shanghai Key Laboratory for Particle Physics and Cosmology; German Research Foundation DFG under Contracts Nos. 455635585, FOR5327, GRK 2149; Istituto Nazionale di Fisica Nucleare, Italy; Ministry of Development of Turkey under Contract No. DPT2006K-120470; National Research Foundation of Korea under Contract No. NRF-2022R1A2C1092335; National Science and Technology fund of Mongolia; National Science Research and Innovation Fund (NSRF) via the Program Management Unit for Human Resources \& Institutional Development, Research and Innovation of Thailand under Contract No. B16F640076; Polish National Science Centre under Contract No. 2019/35/O/ST2/02907; The Swedish Research Council; U. S. Department of Energy under Contract No. DE-FG02-05ER41374.


\bibliographystyle{unsrt}

\begin{thebibliography}{**}

\bibitem{Belle:2003nnu}
S.~K.~Choi \textit{et al.} (Belle Collaboration),
Phys. Rev. Lett. \textbf{91}, 262001 (2003).

\bibitem{babar-y4260-1}
B.~Aubert {\it et al.}  (BaBar Collaboration),
Phys.\ Rev.\ Lett.\  {\bf 95}, 142001 (2005).

\bibitem{ref_tetra_bes1}
M.~Ablikim \textit{et al.} (BESIII Collaboration),
Phys. Rev. Lett. \textbf{110}, 252001 (2013).

\bibitem{ref_tetra_belle}
Z.~Q.~Liu \textit{et al.} (Belle Collaboration),
Phys. Rev. Lett. \textbf{110}, 252002 (2013)
[erratum: Phys. Rev. Lett. \textbf{111}, 019901 (2013)].


\bibitem{bes-ppjpsi}
M.~Ablikim \textit{et al.} (BESIII Collaboration),
Phys. Rev. Lett. \textbf{118}, 092001 (2017).

\bibitem{bes-pphc}
M.~Ablikim \textit{et al.} (BESIII Collaboration),
Phys. Rev. Lett. \textbf{118}, 092002 (2017).

\bibitem{bes-pppsi}
M.~Ablikim \textit{et al.} (BESIII Collaboration),
Phys. Rev. D \textbf{104}, 052012 (2021).

\bibitem{bes-open-xp}
M.~Ablikim \textit{et al.} (BESIII Collaboration),
JHEP \textbf{05}, 155 (2022).

\bibitem{bes-open-wb}
M.~Ablikim \textit{et al.} (BESIII Collaboration),
Phys. Rev. Lett. \textbf{122}, 102002 (2019).

\bibitem{bes-open-ddp}
M.~Ablikim \textit{et al.} (BESIII Collaboration),
Phys. Rev. Lett. \textbf{130}, 121901 (2023).

\bibitem{belle-kkjpsi}
C.~P.~Shen \textit{et al.} (Belle Collaboration),
Phys. Rev. D \textbf{89}, 0720015 (2014).

\bibitem{bes-kkjpsi1}
M.~Ablikim \textit{et al.} (BESIII Collaboration),
Phys. Rev. D \textbf{97}, 071101 (2018).

\bibitem{bes-kkjpsi2}
M.~Ablikim \textit{et al.} (BESIII Collaboration),
Chin. Phys. C \textbf{46}, no.11, 111002 (2022).

\bibitem{bes-kkjpsi3}
M.~Ablikim \textit{et al.} (BESIII Collaboration),
Phys. Rev. Lett. \textbf{131}, 211902 (2023).


\bibitem{bes-dds}
M.~Ablikim \textit{et al.} (BESIII Collaboration),
Phys. Rev. Lett. \textbf{131}, 151903 (2023).

\bibitem{b_psipi1}
S.~K.~Choi \textit{et al.} (Belle Collaboration),
Phys. Rev. Lett. \textbf{100}, 142001 (2008).

\bibitem{bes_psipi}
M.~Ablikim \textit{et al.} (BESIII Collaboration),
Phys. Rev. D \textbf{96}, 032004 (2017)
[erratum: Phys. Rev. D \textbf{99}, 019903 (2019)].



\bibitem{ref_tetra_zcs}
M.~Ablikim \textit{et al.} (BESIII Collaboration),
Phys. Rev. Lett. \textbf{126}, 102001 (2021).


\bibitem{ref_tetra_zcs_lhcb}
R.~Aaij \textit{et al.} (LHCb Collaboration),
Phys. Rev. Lett. \textbf{127}, 082001 (2021).


\bibitem{zcs_theory}
J.~Ferretti and E.~Santopinto,
JHEP \textbf{04}, 119 (2020).

\bibitem{zcs_ar}
J.~Ferretti and E.~Santopinto,
Sci. Bull. \textbf{67}, 1209 (2022).





\bibitem{BES} M.~Ablikim {\it et al.}  (BESIII Collaboration), Nucl.\ Instrum.\ Meth.\ A {\bf 614}, 345 (2010).

\bibitem{KKMC} S.~Jadach, B.~F.~L.~Ward and Z.~Was, Comp.\ Phys.\ Commu. {\bf 130}, 260 (2000); Phys. Rev. D {\bf 63}, 113009 (2001).

\bibitem{evtgen} D.~J.~Lange, Nucl.\ Instrum.\ Meth.\ A {\bf 462}, 152 (2001).

\bibitem{geant} S.~Agostinelli {\it et al.} (GEANT4 Collaboration), Nucl.\ Instrum.\ Meth.\ A {\bf 506}, 250 (2003).



\bibitem{PDG2022}
R.~L.~Workman \textit{et al.} (Particle Data Group),
PTEP, 2022, 083C01 (2022).

\bibitem{supp}
See Supplemental Material  
for the cross section measurement, the significance, and
the systematic uncertainty at each energy point.




\bibitem{QED-Delt}
E.~A.~Kuraev and V.~S.~Fadin, Yad.~Fiz. {\bf 41}, 733 (1985).


\bibitem{isr_med}
W.~Sun, T.~Liu, M.~Jing, L.~Wang, B.~Zhong and W.~Song,
Front. Phys. (Beijing) \textbf{16}, 64501 (2021).


\bibitem{Vacuum-Delt}
S.~Actis {\it et al.}, Eur.~Phys.~J.~C {\bf 66}, 585 (2010).




\bibitem{bes_ppj}
M.~Ablikim \textit{et al.} (BESIII Collaboration),
Phys. Rev. Lett. \textbf{118}, 092001 (2017).

\bibitem{babar_ppj}
J.~P.~Lees \textit{et al.} (BaBar Collaboration),
Phys. Rev. D \textbf{86}, 051102 (2012).

\bibitem{trk}
M.~Ablikim \textit{et al.} (BESIII Collaboration),
Phys. Rev. Lett. \textbf{118}, 092001 (2017).

\bibitem{pid}
M.~Ablikim \textit{et al.} (BESIII Collaboration),
Phys. Rev. Lett. \textbf{117}, 232002 (2016).

\bibitem{lum}
M.~Ablikim \textit{et al.} (BESIII Collaboration),
Chin. Phys. C \textbf{46}, 113003 (2022).



\end{thebibliography}

\end{document}